# Security modeling and efficient computation offloading for service workflow in mobile edge computing

Binbin Huang[a], Zhongjin Li[a*], Peng Tang[b], Shangguang Wang[b], Jun Zhao[c],

Haiyang Hu[a], Wanqing Li[a], Victor Chang[d]

[a]*School of Computer, Hangzhou Dianzi University, Hangzhou, China, 310018, huangbinbin@hdu.edu.cn*

[b]*State Key Laboratory of Networking and Switching Technology, Beijing University of Posts and Telecommunications, Beijing, China*

[c]*School of Computer Science & Engineering, Nanyang Technological University, Singapore,*

[d]*International Business School Suzhou, Xi'an Jiaotong Liverpool University, Suzhou, China*

**Abstract**

It is a big challenge for resource-limited mobile devices (MDs) to execute various complex and energy-consumed mobile applications. Fortunately, as a novel computing paradigm, edge computing (MEC) can provide abundant computing resources to execute all or parts of the tasks of MDs and thereby can greatly reduce the energy of MD and improve the QoS of applications. However, offloading workflow tasks to the MEC servers are liable to external security threats (e.g., snooping, alteration). In this paper, we propose a security and energy efficient computation offloading (SEECO) strategy for service workflows in MEC environment, the goal of which is to optimize the energy consumption under the risk probability and deadline constraints. First, we build a security overhead model to measure the execution time of security services. Then, we formulate the computation offloading problem by incorporating the security, energy consumption and execution time of workflow application. Finally, based on the genetic algorithm (GA), the corresponding coding strategies of SEECO are devised by considering tasks execution order and location and security services selection. Extensive experiments with the variety of workflow parameters demonstrate that SEECO strategy can achieve the security and energy efficiency for the mobile applications.

## 1. Introduction

Recently, MDs (e.g., smart phones and tablets) have become an integral part of our lives due to their portability and compactness. For a single MD, there may be various mobile applications executing on it, such as virtual reality (VR) and face recognition [1-4]. To process these complex mobile applications efficiently, it requires MDs to be resources-riched ( i.e., high computing capacity and battery power) [2, 3]. Unfortunately, MDs are usually resource-constrained due to their physical size. The conflict between the ever-growing resource requirements of mobile applications and the limited resource capacity of MDs impose a big challenge for mobile application execution and drives the transformation of computing paradigm [5].

Many mobile applications, such as image process applications and augmented reality (AR) applications [6], are typical workflow models. Generally, a workflow is composed of multiple procedures/components, and it can be partitioned into a sequence of precedence-constrained tasks [7, 8]. Due to insufficient MD resource, it is impractical to execute complex and energy consuming applications on MD. To address this problem, MDs can offload all or partial tasks of workflow to the cloud in mobile cloud computing. However, since MDs are logically and spatially distant from cloud servers, the bandwidth between cloud servers and MDs have very limited connectivity, which leads to huge communication latency.

Mobile edge computing (MEC) has emerged as a solution to limitations of mobile cloud computing. Fig. 1 shows the architecture of MEC, which mainly includes evolved NodeBs (eNB) and MDs, where eNBs represent network edge equipments (e.g., wireless access points (APs) or base stations) with enormous computation and storage resources. These eNBs can provide computing services to MDs. Since eNBs are in close proximity to MDs, MDs can offload tasks to eNBs directly through pervasive wireless access network, thereby which can significantly reduce the transmission latency [9, 10]. Hence, it is very appropriate to offload partial computation tasks of workflows to MEC servers, which can greatly reduce MDs' energy consumption.

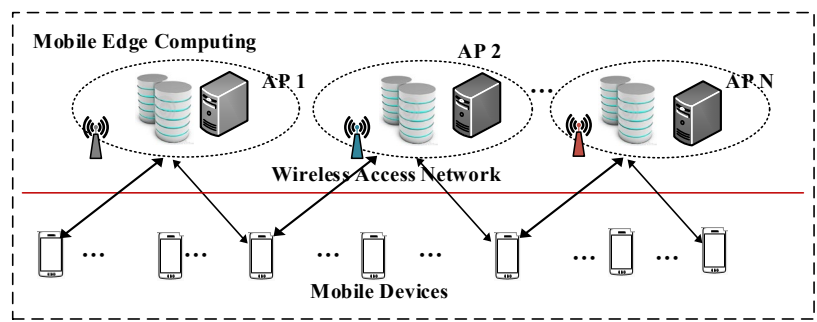

Fig.1. The architecture of computation offloading in mobile edge computing

In addition to the MD's energy consumption, security is another critical concern for mobile applications on cloud computing [11-17], mobile cloud computing [18-20] and mobile edge computing [21-25]. A recent survey reveals that one of the top concerns is security in mobile edge computing [21-26]. In particular, these tasks offloaded to the edge servers are vulnerable to hostile attacks from outside. For example, the information passing between eNBs and MDs can be tampered from hostile actors. However, to the best of our knowledge, few researches consider the security problem of workflow scheduling in MEC. Hence, it is an urgent need to employ the security service to ensure the safety of the security-critical workflow applications in MEC. However, using security services inevitably incurs lots of extra computation time overhead, which will increase energy consumption of MD and the makespan of workflows.

To meet the aforementioned challenges, we propose a security and energy efficient computation offloading (SEECO) strategy for service workflows in MEC environment, the goal of which is to optimize the energy consumption under the risk probability and deadline constraints. First, in order to measure the execution time of security services (i.e., integrity service and confidentiality service), we model the security services overhead under different performance parameters, such as the CPU cores and computation frequency of MEC servers and the size of protected dataset. Then, we take into account the MD's energy consumption, the security requirement and deadline of workflow application, and formulate the security and energy efficient computation offloading problem. Finally, since this problem is NP-hard, a SEECO strategy based on the genetic algorithm (GA) is proposed, and the corresponding coding strategies of which are devised by considering tasks execution order and location and security services selection. Extensive experimental results and analysis demonstrate that SEECO strategy can minimize MD's energy consumption under the risk probability and deadline constraints. In conclusion, the main contributions of this paper can be summarized as follows:

- We build a security overhead model which takes the influence of different performance parameters into account, such as the size of protected dataset, the CPU cores, computation frequency of MEC servers.
- We mainly focus on the computation offloading for workflow type mobile applications, which is much more complex in comparison to the ones with independent tasks.
- We propose a SEECO strategy to minimize the MD's energy consumption under the risk probability and deadline constraints. In particular, SEECO strategy can achieve the security guard for the security-critical tasks in MEC.

We organize this paper as follows. Section 2 summarizes the related work. Section 3 builds a security overhead model that is used to measure the quantity value of security overheads. Section 4 describes problem formulation. Section 5 presents a SEECO strategy for workflow applications to minimize the MD's energy under deadline and security constraints. Section 6 describes the experimental setup and analyzes experimental results. Section 7 concludes this paper and identify future directions.

## 2. Related work

There exist lots of work on workflow scheduling problem in the cloud and mobile cloud computing. In particular, in [27], an evolutionary multi-objective optimization (EMO)-based algorithm is proposed to minimize the makespan and execution cost of workflow in the cloud. In [28], a particle swarm optimization (PSO)-based algorithm is introduced to optimize the execution cost of workflow under deadline constraints. In [29], a Heterogeneous Budget Constrained Scheduling (HBCS) algorithm is designed to minimize the workflow execution time. In [30], MOHEFT is proposed to schedule workflows in Amazon EC2. In [31], using cloud-based computing resources, some analytical models are constructed to quantify the network performance of scientific workflows, and a task scheduling problem to minimize the makespan while meeting a user-specific budget constraint is formulated. In [32], a budget-aware workflow scheduling method is presented in cloud computing environment. However, these scheduling methods above don't take into account the security problem for workflow applications.

With the escalation of the security threatens of data in the cloud or mobile cloud environments, some measures have been implemented to protect security-critical applications. Specifically, in [16] a task-scheduling framework with three feature is presented for security sensitive workflow framework. In [17], a SCAS scheduling scheme is proposed to optimize the workflow execution cost under the makespan and security constraints in clouds. In [11], a SABA scheduling scheme is designed to minimize the makespan under the security and budget constraints. In [33], a security-aware workflow scheduling framework is designed to minimize the makespan and execution cost of workflow while meeting the security requirement. However, to the best of our knowledge, all methods above are mainly designed for the workflow scheduling in cloud computing or mobile cloud computing environment. They are not suitable for workflow scheduling in MEC.

As an emerging paradigm, MEC has attracted considerable attention in the literature [9, 34]. Some works considering computation offloading for MEC have been done, which can be divided into three categories: (i) latency based computation offloading [35-38], (ii) energy based computation offloading [39, 40] and (iii) energy and latency based computation offloading [41-46].

For latency based computation offloading, the objective is to reduce the execution time of mobile applications. Specifically, in [35], a dynamic offloading strategy is proposed to minimize the makespan of mobile applications. In [36], an offline heuristic approach is designed to optimize the average makespan of all users. In [37], a heuristic load-balancing program-partitioning algorithm is proposed. In [38], a polynomial-time approximate algorithm is presented to guarantee performance.

For energy-based computation offloading, the objective is to reduce the MD's energy consumption by offloading computation tasks to edge servers. In particular, in [39] a joint optimization framework is proposed for the radio and computational resource usage by both considering energy consumption and latency. In [40], an OFDMA (time-division multiple access and orthogonal frequency-division multiple access) scheme is designed to minimize the multiple MD's energy consumption.

For energy and latency based computation offloading, the objective is to optimize the MD's energy consumption and the execution time of mobile applications. In particular, in [41], some general guidelines are proposed to minimize the energy consumption and execution time. In [42], a locally optimal algorithm is proposed to optimize the MD's energy and latency. In [43], an algorithmic is designed and implemented using graph theory. In [44], a semi-mobile devices platform framework is proposed to minimize the energy consumption and execution time. In [45], a Lyapunov optimization-based algorithm is introduced to optimize the execution energy and latency. In [46], another Lyapunov optimization-based algorithm is proposed for cloud offloading scheduling and cloud execution output download scheduling.

However, none of the above work considers the impact of task dependency on computation offloading and the security issue for mobile applications. In fact, many mobile applications consist of multiple processes/components (for example, computing components in AR applications), and dependencies between different processes/components cannot be ignored. Because it greatly affects the offloading process. In addition, security cannot be ignored, because it is a key issue in MEC. Therefore, the above schemes are not suitable for security-aware workflow scheduling in MEC. In this paper, we mainly focus on security awareness and energy-efficient workflow scheduling in MEC. We try to minimize the MD's energy consumption under the risk probability and deadline constraints.

## 3. Security Overhead Model

Various safety threats are escalating. Not surprisingly, one of the top concerns is security in mobile edge computing environment [24, 25, 47-49]. Malicious attacks greatly diminish the benefits of mobile edge computing. Hence, it is urgent to need employ various types of security services to protect security-critical workflow application executing in mobile edge computing from malicious attacks. There are three different types of malicious attacks, such as snooping, alteration, and spoofing. To protect the workflow applications against these attacks, three security services, such as authentication service, integrity service, and confidentiality service, can be flexibly selected to form an integrated security protection.

Since security services incur security overheads and the security overhead is node dependent, it is critical and fundamental to measure the quantity value of security overheads for multi-level security service on heterogeneous edge servers. Unfortunately, existing security overhead models [16, 50, 51], only take into account the relationship between the amount of data to be protected and the security overheads with a given number of processor cores and processor frequencies, which are not sophisticated enough yet to consider the node heterogeneity problem. To address this issue, we explore the relationship between the number of processor cores, the processor frequency, the secured data size and the security overheads. And we build an effective security overhead model to approximately measure the security overheads. According to the security overhead model, schedulers enable to incorporate security overheads into workflow scheduling problem.

Since the security overhead of authentication service is a constant value and very small, it usually can be negligible [16]. To examine the security overhead incurred by tasks on heterogeneous edge servers, we test confidentiality service and integrity service, respectively. According to the experiment data, we build a quantitative model to measure the relationship between the security overhead and the secured data size, the number of processor cores, the processor frequency.

*3.1 The computation of security levels*

This section mainly illustrates how to compute the cryptographic speed and the security level according to the security overhead of two security services, respectively. For the sake of simplicity, confidentiality service and integrity service can be represented by $cf$ and $ig$, respectively.

The cryptographic algorithm sets for confidentiality service and integrity services are denoted as $CI_j = \{ci_j^1, ci_j^2, \ldots, ci_j^l, \ldots ci_j^{N(j)}\}$, $j \in \{cf, ig\}$, where $N(j)$ represent the count of cryptographic algorithms for $j$th security service, $ci_j^l$ represents the $l$th cryptographic algorithm of the $j$th security service. A certain cryptographic algorithm $ci_j^l \in CI_j$ can be denoted as a triple $\langle sl(ci_j^l), sp(ci_j^l), cost(ci_j^l, cpu_j^k, f_j^k, \alpha_i) \rangle$, where $sl(ci_j^l)$ represents the security level of $ci_j^l$, $sp(ci_j^l)$ represents the cryptographic speed of $ci_j^l$, and $cost(ci_j^l, cpu_j^k, f_j^k, \alpha_i)$ represents the security overheads of tasks $t_i$ with security level $sl(ci_j^l)$ on edge server $vm_j^k$, respectively. Moreover, $\alpha_i$ represents the secured data size (in bits) of task $t_i$, $cpu_j^k$ represents the number of processor cores of edge server $vm_j^k$, and $f_j^k$ represents the processor frequency of edge server $vm_j^k$, respectively.

For aforementioned cryptographic algorithms, their computational overheads are measured on a Dell R530 server, who is configured with one CPU (2.2GHz 8 Core). In the case of a single core 2.2GHz CPU, it performs these cryptographic algorithms for 100 megabytes (MB) of data. Table 1 shows the security overheads of confidential service, and Table 2 shows that of integrity service.

The fifth column of Table 1 and Table 2 respectively show the security overheads for five encryption algorithms of confidentiality service and five hash functions of integrity service. Based on the experimental data for the security overhead, the cryptographic speed (MB/s) for these cryptographic algorithms can be calculated, and are shown in the fourth column of Table 1 and Table 2. Similar to [16, 50, 51], the security level of these cryptographic algorithms is normalized in a range from 0 to 1. According to the cryptographic speed, the strongest yet slowest encryption algorithm is assigned the security level 1, and then the security level for the rest of the cryptographic algorithm can be calculated.

For example, we use the confidentiality service to show how we calculate the encryption speed and the security level for each security algorithm according to the computation overhead.

First, the encryption speed can be computed by Eq. (1).

$$sp(ci_j^l) = \alpha_i / cost(ci_j^l, cpu_j^k, f_j^k, \alpha_i), j \in \{cf, ig\}, 1 \leq l \leq 5. \quad (1)$$

where $\alpha_i = 100$ $cpu_j^k = 1$ $f_j^k = 2.2$. And then the strongest yet slowest encryption algorithm, IDEA (see Table 1) is assigned the security level 1. Security levels of the encryption algorithms are proportional to their computation overhead. Hence, security levels for the rest of the encryption algorithms can be computed by Eq. (2).

$$sl(ci_j^l) = cost(ci_j^l, cpu_j^k, f_j^k, \alpha_i) / cost(ci_{cf}^1, cpu_j^k, f_j^k, \alpha_i), j \in \{cf, ig\}, 1 \leq l \leq 5. \quad (2)$$

where $\alpha_i = 100$ $cpu_j^k = 1$ $f_j^k = 2.2$. Similarly, the computation overhead for the integrity service is listed in Table 2. In accordance with the computation overhead, the hash speed can be computed by Eq.

(1). According to the hash speed, the strongest yet slowest hash function Tiger is assigned the security level 1, and the security levels for the other hash functions can be computed by Eq. (2).

Table 1. The encryption algorithms for confidential service

| Symbols $ci_{cf}^l$ | Encryption Algorithms | Level $sl(ci_{cf}^l)$ | Speed(Mb/s) $sp(ci_{cf}^l)$ | Computation Overhead $cost(ci_{cf}^l)$(s) |
|---|---|---|---|---|
| $ci_{cf}^1$ | IDEA | 1.0 | 11.76 | 8.50 |
| $ci_{cf}^2$ | DES | 0.85 | 13.83 | 7.23 |
| $ci_{cf}^3$ | AES | 0.53 | 22.03 | 4.54 |
| $ci_{cf}^4$ | Blowfish | 0.56 | 20.87 | 4.79 |
| $ci_{cf}^5$ | RC4 | 0.32 | 37.17 | 2.69 |

Table 2. The hash functions for integrity service

| Symbols $ci_{ig}^l$ | Hash Functions | Level $sl(ci_{ig}^l)$ | Speed(Mb/s) $sp(ci_{ig}^l)$ | Computation Overhead $cost(ci_{cf}^l)$(s) |
|---|---|---|---|---|
| $ci_{ig}^1$ | TIGER | 1.0 | 75.76 | 1.32 |
| $ci_{ig}^2$ | RipeMD160 | 0.75 | 101.01 | 0.99 |
| $ci_{ig}^3$ | SHA-1 | 0.69 | 109.89 | 0.91 |
| $ci_{ig}^4$ | RipeMD128 | 0.63 | 119.05 | 0.94 |
| $ci_{ig}^5$ | MD5 | 0.44 | 172.41 | 0.58 |

*3.2 The computational overheads for the secured data size*

In this section, we explore the influence of the secured data size on the security overhead. We tested the different secured data size on a Dell R530 server with a single core 2.2GHz CPU. The mean size of the secured data varies from 100 MB to 1000 MB. Fig. 2(a) shows the security overheads for five encryption algorithms of confidential service, and Fig. 2(b) shows the security overheads for five hash functions of integrity service.

From Fig. 2(a), we can observe two important features. First, with the secured data size increasing, the computational overheads for these five cryptographic algorithms (IDEA, DES, AES, Blowfish and RC4) increase linearly. Second, when the size of secured data is constant, the relationship of the computational overhead for these five cryptographic algorithms: IDEA>DES> Blowfish>AES>RC4. The computational overhead of the encryption service experienced by different size of secured data can be computed by Eq. (3).

$$cost(ci_j^l, cpu_j^k, f_j^k, \alpha_j) = \alpha_j * cost(ci_j^l, cpu_j^k, \alpha_i)/\alpha_i, j \in \{cf, ig\}, 1 \leq l \leq 5. \quad (3)$$

where $\alpha_j$ represents the size of secured data, $\alpha_i = 100$, $cpu_j^k = 1$ and $f_j^k = 2.2$, $cost(ci_j^l, cpu_j^k, f_j^k, \alpha_i)$ represents the computational overhead experienced by the 100 megabytes of data with security level requirements $ci_j^l$ on an edge server with a single core 2.2GHz CPU.

Fig. 2(b) shows the experimental results for these five hash algorithms. We observe from Fig. 2(b) that the security overheads of these five hash algorithms (TIGER, RipeMD160, SHA-1, RipeMD128 and MD5) increase linearly with the secured data size increasing. Moreover, when the secured data size is constant, the relationship of the computational overhead for these five hash algorithms: TIGER >RipeMD160>SHA-1>RipeMD128 >MD5. The computational overhead of the five hash algorithms experienced by different size of secured data can be computed by Eq. (3).

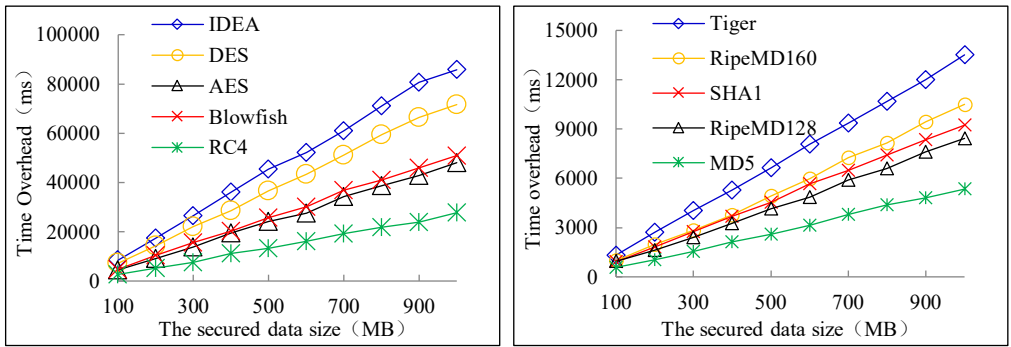

(a) The relationship between data size and the security overhead

(b) The relationship between data size and the integrity overhead

Fig. 2. The security overhead with different data size

*3.3 The computational overheads for the number of processor cores*

To examine the influence of different processor cores on the security overhead, in the set of experiments, the processor cores are varied from 1 to 8 with increments of 1. We measured the quantity value of security overheads experienced with 100M security-required data on a Dell R530 server with a 2.2GHz CPU. The security overheads of confidential service obtained with experiments are shown in Fig. 3(a), and that of hash functions for integrity service are shown in Fig. 3(b).

Fig. 3(a) shows that when the processor frequency and the secured data size are constant, the computational overheads for these five cryptographic algorithms (IDEA, DES, AES, Blowfish and RC4) decrease with the number of processor cores increasing. On the other hand, when the number of processor cores and the processor frequency are constant, the relationship of the security overhead for these five cryptographic algorithms experienced by the same data size: IDEA>DES> Blowfish>AES>RC4. Therefore, when the processor frequency and the amount of data are constant, the security overheads for the encryption algorithms are inversely proportional to the number of processor cores, which can be computed by Eq. (4).

$$cost(ci_j^l, cpu_j^k, f_j^k, \alpha_j) = cost(ci_j^l, cpu_j^k, f_j^k, \alpha_i)/cpu_j^k, j \in \{cf, ig\}, 1 \leq l \leq 5. \tag{4}$$

where $cpu_j^k = 1$, $f_j^k = 2.2$, $cost(ci_j^l, 1, 2.2, \alpha_i)$ represents the computational overhead experienced by the $\alpha_i$ megabytes of data with security level requirements $ci_j^l$ on an edge server with a single core 2.2GHz CPU.

Fig. 3(b) shows the experimental results for these five hash algorithms. We observe from Fig. 3(b) that when the processor frequency and the secured data size are constant, the computational overheads for these five hash algorithms (TIGER, RipeMD160, SHA-1, RipeMD128 and MD5) decrease with the increased number of processor cores. Moreover, when the number of processor cores and the processor frequency are constant, the relationship of the computational overhead for these five hash algorithms experienced by the same size data: TIGER >RipeMD160> SHA-1>RipeMD128 >MD5. Therefore, the computational overheads of the hash algorithms are inversely proportional to the number of processor cores when the processor frequency and the amount of data are constant, which can be computed by Eq. (4).

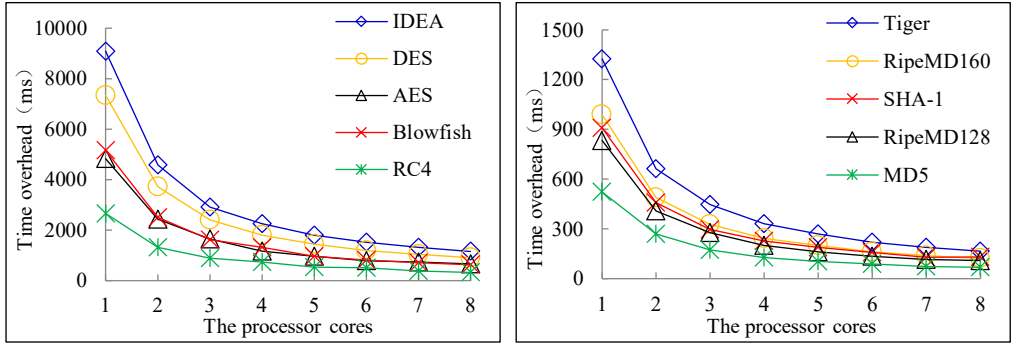

(a) The relationship between processor cores and the security overhead

(b) The relationship between processor cores and the integrity overhead

Fig. 3. The security overhead with different processor cores

*3.4 The computational overhead for the processor frequency*

To examine the influence of different processor frequency on the security overhead, in the set of experiments, the processor frequency is varied from 30% to 100% with increments of 10%. We measured the quantity value of security overheads experienced with 100M security-required data on a Dell R530 server with a single core. The security overheads for confidential service obtained with experiments are shown in Fig. 4(a), and that of hash functions for integrity service are shown in Fig. 4(b).

Fig. 4(a) shows the experimental results for the computational overhead of these five cryptographic algorithms. We observe from Fig. 4(a) that when the processor cores and the secured size are constant, the computational overheads for these five cryptographic algorithms (IDEA, DES, AES, Blowfish and RC4) decrease as the processor frequency increases. On the other hand, when the number of processor cores and the processor frequency are constant, the relationship of the computational overhead for these five cryptographic algorithms experienced by the same data size: IDEA>DES> Blowfish>AES>RC4. Therefore, when the processor cores and the secured data size are constant, the computational overheads of the encryption algorithms are inversely proportional to the processor frequency, which can be computed by Eq. (5).

$$cost(ci_j^l, cpu_j^k, f_j^k, \alpha_j) = cost(ci_j^l, cpu_j^k, F, \alpha_i) * F/f_j^k, j \in \{cf, ig\}, 1 \leq l \leq 5. \qquad (5)$$

where $cpu_j^k = 1$, $\alpha_i = 100$, $F$ is the maximum operating frequency of the processor, and $cost(ci_j^l, 1, F, 100)$ represents the computational overhead experienced by the 100 megabytes of data with security level requirements $ci_j^l$ on an edge server with a single core $F$ GHz CPU.

Fig. 4(b) shows the experimental results for these five hash algorithms. We observe from Fig. 4(b) that when the processor cores and the secured data size are constant, the computational overheads for these five hash algorithms (TIGER, RipeMD160, SHA-1, RipeMD128 and MD5) decrease as the processor frequency increases. Moreover, when the number of processor cores and the processor frequency are constant, the relationship of the computational overhead for these five hash algorithms experienced by the same size data: TIGER >RipeMD160> SHA-1>RipeMD128 >MD5. Therefore, the computational overheads of the hash algorithms are inversely proportional to the processor frequency when the number of processor cores and the amount of data are constant, which can be computed by Eq. (5).

In conclusion, the computation overhead $cost(ci_j^l, cpu_j^k, f_j^k, \alpha_i)$ mainly depends on the cryptographic algorithms used $ci_j^l$, the secured data size $\alpha_i$, the number of processor cores $cpu_j^k$ and the processor frequency $f_j^k$ of the heterogeneous node $vm_j^k$, which can be calculated by Eq. (6):

$$cost(ci_j^l, cpu_j^k, f_j^k, \alpha_i) = (\alpha_i * 2.2)/(speed(ci_j^l) * f_j^k * cpu_j^k), j \in \{cf, ig\}, 1 \leq l \leq 5. \quad (6)$$

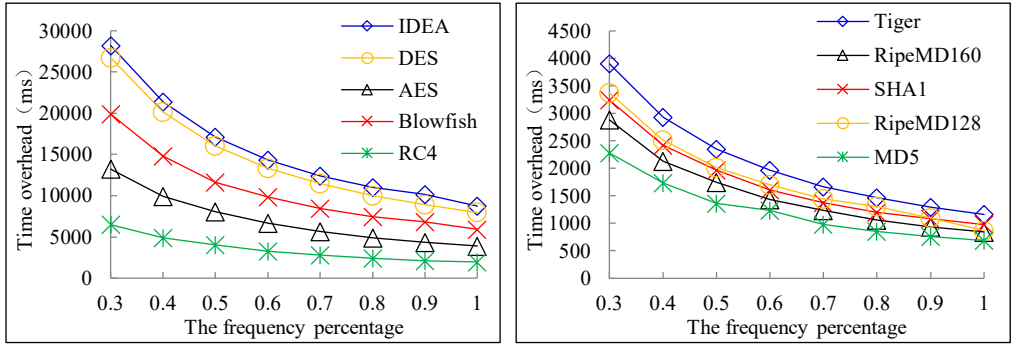

(a) The relationship between processor frequencies and the security overhead

(b) The relationship between processor frequencies and the integrity overhead

Fig. 4. The security overhead with different processor frequencies

## 4. Problem Formulation

In this section, we first introduce a security-aware workflow model and a mobile edge computing model, respectively. Then we analyze the process of security-aware task execution. Next we analyze the energy consumption and risk probability for workflow, respectively. At last, we formulate the security-aware and efficient-energy workflow scheduling problem. To improve the readability, we summarize the notations used in throughout this paper in Table 3.

Table 3. Notations

| Symbols | Definition |
|---|---|
| $W$ | The workflow model; |
| $T$ | The task set that compose workflow; |
| $E$ | The directed edges set; |
| $T_D$ | The deadline for workflow; |
| $P_T$ | The risk probability constraint for workflow; |
| $t_i$ | Task $t_i$ of workflow; |
| $e(i, j)$ | A directed edge; |
| $\alpha_i$ | The input data size of task $t_i$; |
| $\beta_i$ | The output data size of task $t_i$; |
| $\omega_i$ | The workload of task $t_i$; |
| $pre(t_i)$ | The predecessor set of task $t_i$; |
| $succ(t_i)$ | The successor set of task $t_i$; |
| $AP_j$ | The $j$th wireless access point; |
| $B$ | The communication bandwidth between any two APs; |
| $vm_j^k$ | The $k$th virtual machine in the $j$th wireless AP; |
| $f_j^k$ | The processor frequency of $vm_j^k$; |
| $cpu_j^k$ | The number of processor cores of $vm_j^k$; |
| $capability_j^k$ | The processor capability of $vm_j^k$; |
| $P_{01}^{Comp}$ | The MD's computation power; |
| $P_{01}^{UL}$ | The MD's transmitting power; |
| $P_{01}^{DL}$ | The MD's receiving power; |
| $B_{jk}^{UL}$ | The uplink channel bandwidths between $vm_j^k$ and MD; |
| $B_{jk}^{DL}$ | The downlink channel bandwidths between $vm_j^k$ and MD; |
| $T^{TR}(t_{i-1})$ | The transfer time of output data of task $t_{i-1}$; |
| $ECost(t_{i-1})$ | The total security overhead of cryptographic service of task $t_{i-1}$; |
| $DECost(t_i)$ | The security overheads of all of the immediate processors tasks of task $t_i$; |
| $T^{EX}(t_i, vm_n^q)$ | The execution time $T^{EX}(t_i, vm_n^q)$ of task $t_i$ on $vm_n^q$; |
| $T^{PR}(t_i, vm_n^q)$ | The total processing time $T^{PR}(t_i, vm_n^q)$ of task $t_i$ on VM $vm_n^q$; |
| $E^{Comp}(t_i)$ | The MD's computation energy consumption; |
| $E^{UL}(t_i)$ | The MD's upload energy consumption; |
| $E^{DL}(t_i)$ | The MD's download energy consumption; |
| $P(t_i, sl(ci_j^l))$ | The risk probability of the security service $ci_j^l$ of task $t_i$; |
| $P(t_i)$ | The risk probability of task $t_i$; |
| $P(W)$ | The risk probability of workflow; |
| $T^{ST}(t_i)$ | The start time of task $t_i$; |
| $T^{ET}(t_i)$ | The end time of task $t_i$; |
| $T(W)$ | The total execution time of workflow; |

*4.1 Security-aware workflow model*

A security-aware workflow model can be represented by a four-dimensional tuple $W = (T, E, T_D, P_T)$. $T = \{t_0, t_1, ..., t_i, ..., t_{n-1}\}$ denotes the set of $n$ tasks. Each task $t_i$ can be represented by a tuple $\{\alpha_i, \beta_i, \omega_i\}$, in which $\alpha_i$ is the input data size (in bits) of task $t_i$, $\beta_i$ is the output data size of task $t_i$, and $\omega_i$ is the workload of task $t_i$, respectively. $E$ is the directed edge set. A directed edge $e(i,j) \in E$ indicates that task $t_i$ is the predecessor of task $t_j$. It means that task $t_j$ can start being executed only that its predecessor tasks $t_i$ complements. $pre(t_i)$ denotes the predecessor set of tasks $t_i$. $pre(t_i)$ denotes the successor set of task $t_i$. $T_D$ denotes the deadline of workflow $W$. $T_D$ is specified by users according to the workflow application performance requirement. $P_T$ denotes the risk probability constraint of workflow $W$. The value of $P_T$ mainly depends on the sensitivity degree of the workflow in edge servers. The lower the risk probability constraint, the higher the sensitivity degree of the data is.

*4.2 Mobile edge computing model*

In mobile edge computing environment, we mainly consider the scenario where a MD can offload partial tasks of the workflow $W$ to the $M$ wireless APs. We denote the set of APs as $AP = \{AP_0, AP_1, AP_2, ..., AP_j, ..., AP_M\}$, where $AP_0$ denotes the MD. For the convenience of computing, we assume that all these APs have the same communication bandwidth B. The communication bandwidth between different virtual machines (VMs) in any of the $M$ wireless access point APs $AP_j$ ($1 \leq j \leq M$) is infinite.

The set of VMs $VM_j$ that are possessed by any of the $M$ wireless access point APs $AP_j$ can be denoted $VM_j = \{vm_j^1, vm_j^2, ..., vm_j^k, ..., vm_j^{K_j}\}$, where $vm_j^k$ represents the $k$th VM in the $j$th wireless AP, and $K_j$ represents the total number of VMs that are possessed by the APs $AP_j$. Each VM $vm_j^k$ has different configurations, such as the number of processor cores, the processor frequency, and processor capability, etc. We use a triple $vm_j^k = \{f_j^k, cpu_j^k, capability_j^k\}$ to represent the VM $vm_j^k (1 \leq j \leq M, 1 \leq k \leq K_j)$, in which $f_j^k$ is the processor frequency of the $vm_j^k$, $cpu_j^k$ is the number of processor cores of the $vm_j^k$, and $capability_j^k$ is the processor capability of the $vm_j^k$, respectively. Especially, when $j = 0$, $AP_0$ denotes the MD. As the MD' processor is seen as a VM in $AP_0$, the value of $K_0$ is set to 1, and the tuple $vm_0^1 = \{f_0^1, cpu_0^1, capability_0^1\}$ denotes the MD' processor itself, in which $f_0^1$ is the processor frequency of the MD, $cpu_0^1$ is the number of processor core of the MD, $capability_0^1$ is the processor capability of the MD, respectively. Moreover, the power of the MD can be represented by a triple $P_{01} = \{P_{01}^{Comp}, P_{01}^{UL}, P_{01}^{DL}\}$, in which $P_{01}^{Comp}$ is the MD's computation power (in Kbps), $P_{01}^{UL}$ is the MD's transmitting power, and $P_{01}^{DL}$ is the MD's receiving power, respectively. All of them are constant. The uplink rates $C_{jk}^{UL}$ between the MD and the $AP_j$ can be computed by Eq. (7), and the downlink rates $C_{jk}^{DL}$ between them can be computed by Eq. (8).

$$C_{jk}^{UL} = B_{jk}^{UL} log_2(1 + \frac{P^{Tx} h_{ijk}^{UL}}{\bar{\omega}_0}). \tag{7}$$

$$C_{jk}^{DL} = B_{jk}^{DL} log_2(1 + \frac{P_{AP} h_{ijk}^{DL}}{\bar{\omega}_0}). \tag{8}$$

where $B_{jk}^{UL}$ is the uplink channel bandwidth, $B_{jk}^{DL}$ is the downlink channel bandwidth; $P^{Tx}$ is the MD's transmission power, and $P_{AP}$ is the APs' transmission powers; $h_{ijk}^{UL}$ is the uplink channel gain, and $h_{ijk}^{DL}$ is the downlink channel gain; $\bar{\omega}_0$ is the white noise power level.

*4.3 A security-aware task execution process analysis*

Fig. 5 illustrate the security-aware task execution process. Task $t_i$ and task $t_{i+1}$ are the immediate successors of task $t_{i-1}$. We assume task $t_i$ and task $t_{i-1}$ are executed on VM $vm_n^q$ and $vm_m^p$, respectively. When task $t_{i-1}$ is finished, the output data $\beta_{i-1}$ of task $t_{i-1}$ is transferred to its successor task $t_i$, and the corresponding transfer time $T^{TR}(t_{i-1})$ can be computed by Eq. (9).

$$T^{TR}(t_{i-1}) = \begin{cases} \beta_{i-1}/C_{nq}^{UL}, \text{the output data of task } t_{i-1} \text{ on MD is transferred to VM } vm_n^q, \\ \beta_{i-1}/C_{mp}^{DL}, \text{the output data of task } t_{i-1} \text{ on VM } vm_m^p \text{ is transferred to MD}, \\ \beta_{i-1}/B, \text{task } t_{i-1} \text{ and } succ(t_i) \text{ are executed on the different APs}, \\ 0, \text{task } t_{i-1} \text{ and } pre(t_i) \text{ are executed on the same VM or APs}. \end{cases} \tag{9}$$

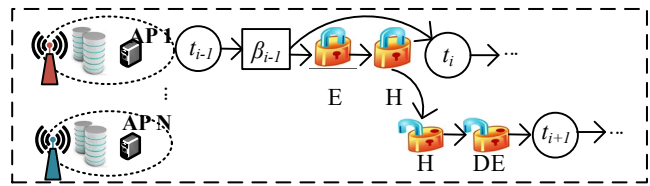

Fig. 5. The task execution process with security services

As Fig. 5 shows, if VM $vm_m^p$ and $vm_n^q$ are on the different APs, the output data $\beta_{i-1}$ of task $t_{i-1}$ need to be transferred to its immediate successor task $t_{i+1}$. Before the output data is transferred, it will be implemented by several security services. Different security services incur distinct computation time overhead. First, in order to protect the output data $\beta_{i-1}$ of task $t_{i-1}$ from snooping attacks, authentication service (denoted as A) is employed to authenticate the user who intends to receive the output data. However, the security overhead of authentication service are negligibly small. And then to protect the output data $\beta_{i-1}$ of task $t_{i-1}$ from spoofing attacks, confidentiality service (denoted as E) is employed to encrypt these data. Next to protect the output data $\beta_{i-1}$ of task $t_{i-1}$ from alteration attacks, integrity service is successively employed to implement a hash algorithm (denoted as H) to them, and the security overhead of encryption service and integrity service are computed by Eq. (6). Hence, the total security overhead of cryptographic service can be computed by Eq. (10). After task $t_{i+1}$ receives the encrypted data from task $t_{i-1}$, the data will be decrypted (denoted as DE) and its integrity will be verified (denoted as IV). Otherwise, if task $t_{i-1}$ and its immediate successor task $t_i$ are executed on the same VM or AP, the output data $\beta_{i-1}$ of task $t_{i-1}$ can be used directly without encrypting. The overall decryption overheads of all of the immediate processors' tasks of task $t_i$ can be computed by Eq. (11).

$$ECost(t_{i-1}) = \sum_{j \in \{cf, ig\}} 2.2 * \beta_{i-1}/(speed(ci_j^l) * f_m^p * cpu_m^p). \tag{10}$$

$$DECost(t_i) = \sum_{t_{i-1} \in pre(t_i)} \sum_{j \in \{cf, ig\}} (cpu_m^p/cpu_n^q) * 2.2 * \beta_{i-1}/(speed(ci_j^l) * f_n^q * cpu_n^q). \tag{11}$$

The task $t_i$ cannot start its execution on a candidate VM $vm_n^q$ util it receives the output data from all of its immediate processors' tasks, and the execution time $T^{EX}(t_i, vm_n^q)$ of task $t_i$ on a candidate VM $vm_n^q$ can be computed by Eq. (12).

$$T^{EX}(t_i, vm_n^q) = \omega_i/capacity_n^q. \tag{12}$$

Based on the aforementioned computation in Eqs (9), (10), (11) and (12), the total processing time $T^{PR}(t_i, vm_n^q)$ of task $t_i$ on a VM $vm_n^q$ can be computed by Eq. (13).

$$T^{PR}(t_i, vm_n^q) = DECost(t_i) + T^{EX}(t_i, vm_j^k) + T^{TR}(t_i) + ECost(t_i). \tag{13}$$

*4.4 Mobile device energy consumption analysis*

The MD's energy consumption mainly consists of the computation energy consumption and wireless transmission energy consumption.

(1) Computational Energy Consumption: when task $t_i$ is executed on the MD, the MD's energy consumption can be computed by Eq. (14).

$$E^{Comp}(t_i) = P_{01}^{Comp} T^{EX}(t_i, vm_j^k), j = 0, k = 1. \tag{14}$$

(2) Wireless Transmission Energy Consumption: the MD's wireless transmission energy consumption $E^{TR}(t_i)$ which consists of the upload energy consumption $E^{UL}(t_i)$ and the download consumption $E^{DL}(t_i)$ can be computed by Eq. (15).

$$E^{TR}(t_i) = E^{UL}(t_i) + E^{DL}(t_i). \tag{15}$$

$$E^{UL}(t_i) = \sum_{t_i, t_s \in T \land t_s \in succ(t_i) \land t_s \text{ on } MEC \land t_i \text{ on } MD} P_{01}^{UL} * \beta_i/C_{nq}^{DL}. \tag{16}$$

$$E^{DL}(t_i) = \sum_{t_i, t_p \in T \land t_p \in pre(t_i) \land t_p \text{ on } MEC \land t_i \text{ on } MD} P_{01}^{DL} * \alpha_i/C_{mp}^{DL}. \tag{17}$$

where $E^{UL}(t_i)$ and $E^{DL}(t_i)$ are the MD's upload and download energy consumption, respectively. If task $t_i$ are executed on the MD and its successor task $t_s$ is executed on the VM $vm_n^q$ which isn't the MD, the output data of task $t_i$ are needed to upload to the VM $vm_n^q$, thereby, producing the upload energy consumption $E^{UL}(t_i)$. Similarly, if task $t_i$ are executed on the MD and its processor task $t_p$ is executed on the VM $vm_m^p$ which isn't the MD, the output data of all of its processor tasks are needed to download to the MD, which incurs the download energy consumption $E^{DL}(t_i)$.

*4.5 The risk probability analysis of workflow*

In MEC environment, the execution of workflow is not risk-free probability, hence, it is important to build the risk probability model to quantitatively calculate the risk probability.

Without loss of generality, we assume that the distribution of risk probability follows a *Poisson* probability distribution for any given time interval. The risk probability $P(t_i, sl(ci_j^l))$ of task $t_i$ is the function of the security level $sl(ci_j^l)$ of the security service $ci_j^l$ employed by task $t_i$, and can be denoted by Eq. (18) [52, 53].

$$P(t_i, sl(ci_j^l)) = 1 - exp(-\lambda_j(1 - sl(ci_j^l))), j \in \{cf, ig\}. \quad (18)$$

In MEC, the risk coefficient $\lambda_j$ is different for encryption service and integrity service. Since 2.5 alteration attacks and 1.8 spoofing attacks are usually suffered in a unit time interval, $\lambda_{cf}$ and $\lambda_{ig}$ are set 2.5 and 1.8, respectively. The risk probability $P(t_i)$ of task $t_i$ which employs these two kinds of security services with different security level can be computed by Eq. (19).

$$P(t_i) = 1 - \prod_{j \in \{cf, ig\}} 1 - P(t_i, sl(ci_j^l)). \quad (19)$$

Given the task set $T$ of the workflow $W$, its risk probability $P(W)$ can be computed by Eq. (20).

$$P(W) = 1 - \prod_{t_i \in T} 1 - P(t_i). \quad (20)$$

As the risk probability constraint of the workflow $W$ is $P_T$, in order to satisfy its security requirement, this comes to the constraint in Eq. (21):

$$P(W) \leq P_T. \quad (21)$$

*4.6 Problem definition*

We focuses on finding one or more feasible solution $\varphi = (Order, Loc, Lev_{cf}, Lev_{ig})$ with minimized MD's energy consumption under the total workflow execution deadline and security constraints. $Order = \{r_0, r_1, \ldots, r_i, \ldots r_{n-1}\}$ is the set of a task execution sequence, in which an index $i$ represents the task execution sequence and its value $r_i$ represents a task whose execution sequence index is $i$; $Loc = \{x_{jk}^i | i \in [0, n-1], j \in [0, M], k \in [1, K_j], M, K_j \in [0, F]\}$ is the set of a task execution location set, where $x_{jk}^i$ is a hexadecimal value, $x_{jk}^i = 0x01$ represents that task $t_i$ is assigned to MD, otherwise, is offloaded to any of the $M$ APs; $Lev_{cf} = \{ci_{cf}(r_i) | r_i \in Order\}$ is the set of a task encryption service level; $Lev_{ig} = \{ci_{ig}(r_i) | r_i \in Order\}$ is the set of a task integrity service level.

To meet with the deadline constraint $T_D$, the total execution time of the workflow $W$ need to be calculated. The execution time of workflow $W$ depends mainly on the finish time of task $t_{n-1}$. The start time and finish time of task $t_i$ can be denoted by represent the $T^{ST}(t_i)$ and $T^{ET}(t_i)$, respectively. The task $t_i$ cannot start to execute until it receives the output data from all of its immediate processors' tasks. This comes to the constraint below:

$$\max_{t_r \in pre(t_i)} \{T^{ET}(t_r) | t_r \in T\} \leq T^{ST}(t_i). \quad (22)$$

$$T^{ET}(t_r) = T^{ST}(t_r) + T^{PR}(t_r, vm_n^q). \quad (23)$$

The total execution time $T(W)$ of the workflow $W$ can be computed by Eq. (24).

$$T(W) = max\{T^{ET}(t_i) | t_i \in T\}. \quad (24)$$

According to the MD's energy consumption analysis as mentioned in 4.4, the total execution energy $E(W)$ of the workflow $W$ can be computed by Eq. (25).

$$E(W) = \sum_{t_i \in T \wedge t_i \text{ on local}} E^{Comp}(t_i) + \sum_{t_i, t_r \in T \wedge t_r \in pre(t_i) \wedge t_r \text{ on MEC} \wedge t_i \text{ on local}} E^{DL}(t_i) + \sum_{t_i, t_f \in T \wedge t_f \in succ(t_i) \wedge t_f \text{ on MEC} \wedge t_i \text{ on local}} E^{UL}(t_i). \quad (25)$$

The primary optimization objective is to find an optimal execution sequence, allocation decision and security service levels for the task set $T$ of the workflow $W$ to minimize the MD's energy consumption under the total workflow execution deadline and risk probability constraints. The constrained optimization problem can be formulated as follows:

$$\text{Minimize: } E(W) \quad (26)$$
$$\text{Subject to: } P(W) \leq P_T, \quad (27)$$

$$T(W) \leq T_D, \qquad (28)$$

where the risk probability constraints of the workflow $W$ can be represented by Eq. (27), and the total workflow execution deadline constraints can be represented by Eq. (28).

**5. SEECO Algorithm Implementation**

The problem to be solved in this paper is NP-hard [54]. Typically, to solve the NP-hard problem, heuristic and meta-heuristic algorithms are usually used. The goal is to find an optimal approximate solution in an acceptable time. Genetic algorithm (GA) which developed by Dr. J. Holland is a meta-heuristic algorithm with reliable global search capability, in which selection, crossover and mutation are used to produce individuals with better fitness. The genetic algorithm doesn't have to calculate the reciprocal or the gradient of the objective function, and it doesn't require that the objective function is continuous, and the algorithm has inherent parallelism and parallel computing ability and the ability of global optimization characteristics, and it's an efficient method for solving optimization problem, and it's widely applied to numerical optimization, assembly optimization, machine learning, image recognition, neural networks, and fuzzy control. Moreover, the genetic algorithm is a robust spatial search technology, which can use the principle of evolution to obtain a feasible solution from a larger search space in linear time. The problem to be solved in this paper is a single-objective constrained optimization problem, which requires to find the approximate optimal solution in a relatively short time. To address this issue, we present a SEECO strategy based on an improved genetic algorithm. The algorithm's process consists of the following steps:

    (1) Encoding the task execution order, task execution location, encryption service level and integrity service level, respectively.

    (2) Generating the initial population randomly for the first generation.

    (3) Generating a new generation of the population by selection, crossover and mutation operators.

    (4) Evaluating each individual in the population by using a fitness function and selecting the individuals with the best fitness value in a new population.

    (5) To continue to iterate until a specified maximum number of iterations is met.

The related implementation steps are introduced in detail in the next sections.

*5.1 Encoding*

To solve the problem, the solution of the problem need to be transformed into the chromosome embodied by code. Here, we first make a topological sort for task execution order in the workflow application $W$, and then assign an integer index to each task according to the sorting results. The index starts from 0. A solution is devised as a four-tuple containing a task execution sequence set $Order = \{r_0, r_1, \ldots, r_i, \ldots r_{n-1}\}$, a task execution location set $Loc = \{loc(r_0), loc(r_1), \ldots, loc(r_i), \ldots, loc(r_{n-1})\}$, a task encryption service level set $Lev_{cf} = \{ci_{cf}(r_0), ci_{cf}(r_1), \ldots, ci_{cf}(r_i), \ldots ci_{cf}(r_{n-1})\}$, and a task integrity service level set $Lev_{ig} = \{ci_{ig}(r_0), ci_{ig}(r_1), \ldots, ci_{ig}(r_i), \ldots ci_{ig}(r_{n-1})\}$. The task execution order set $Order$ is a vector containing a permutation of all tasks, in which an index $i$ denotes the task execution sequence and its value $r_i$ denotes the task whose execution sequence index is $i$. The task execution location set $Loc$ is also vector, in which an index denotes a task execution order and its value represents the VM on which the task corresponding execution order is executed. Similarly, the third set $Lev_{cf}$ and the fourth set $Lev_{ig}$ are two $n$-length vectors, in each which an index represents a task execution order and its value represents the encryption service level and the integrity service level employed by the task corresponding execution order, respectively.

An example of the workflow $W$ is shown in Fig. 6. A valid scheduling order is shown in Fig. 7, and the encoding of a possible schedule for this workflow is given in Fig. 7. Moreover, the mappings from the tasks to the VMs and from the tasks to the two security service levels are also given in Fig. 7.

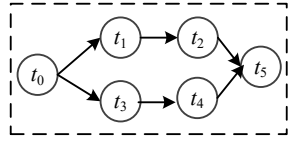

Fig. 6. An example of a workflow

Fig. 7. Encoding scheme of a valid schedule for the workflow

*5.2 Genetic Operators*

*5.2.1 Selection*

In the selection stage, we select chromosome recombination to generate the next population through crossover and mutation. The binary contest selection method is used. In tournament selection, two individuals are randomly selected from the population and compared according to their fitness and the sum of constraint violation. Better solutions are selected and kept in intermediate populations. This process continues until all *N* populations are filled.

In order to deal with these constraints, the superiority of the feasible solution method [55] is adopted, in which a set of three feasible criteria are used: (1) the optimal solution (according to the fitness function) is better of two feasible solutions; (2) the feasible solution is always better than the infeasible solution, (3) the optimal solution has the smaller sum of the constraint violation of two feasible solutions. In this article, we can calculate the sum of the constraint violations as follows:

$$Violate = max(0, T(W) - T_D) + max(P(W) - P_T). \qquad (29)$$

*5.2.2 Crossover*

Crossover operator is the most important genetic operation of genetic algorithm. It refers to the operation in which the partial structure of two parent individuals is replaced and recombined to form new ones. The role of crossover is to generate offspring that are better individuals by preserving partial individuals from the parents. Moreover, it plays the role of searching global and exploring the unknown space. Hence, it finds the better and better solutions by the crossover operator. In this section, according to the coding as mentioned in 5.1, we perform the crossover operator to the set $Order$, the set $Loc$, the $Lev_{cf}$ strings and the $Lev_{ig}$ strings, respectively. The single-point crossover operator procedures for the four different settings are introduce as follow.

**Definition 1**. The match area: the task sequence between the first task to the cut-off position in the sorted tasks set.

A valid scheduling order must meet the precedence-constraint of the tasks in workflow. For example, task $t_j$ is the successor of task $t_i$, $t_j$ cannot start execution until its precedent task $t_i$ complements in a task execution order individual $Order$. The new individual which are generated by crossover operation must also meet these constraints. To meet the precedence-constraint of the tasks in workflow, in reference to the literature [27], the process of the crossover operator for the task execution order $Order$ is shown as Algorithm 1. First, the operator generates at random a number $r \in [0, n-1]$ as a cut-off position, and generates the match area of $Order_1$ and $Order_2$, respectively (Step 3-5). After that, the match area of the individual $Order_1$ is prepended to $Order_2$, the match area of the individual $Order_2$ is prepended to $Order_1$, and two temporary new individuals $Order_{12}$ and $Order_{21}$ are produced (Step 7-8). Then, each temporary new individual is scanned from the beginning, and the repetitive tasks in two temporary new individuals are removed, and get their offsprings (Step 9-10). An example of this operation is given in Fig. 8, in which we choose randomly the task with execution sequence index 1 as the cut-off position. Then, according to the Algorithm 1, it performs the crossover operator on $Order_1$ and $Order_2$.

| Algorithm 1: The single crossover of task execution order |
|---|
| BEGIN |
| 01. Generated at random a number $r \in [0,n-1]$; |
| 02. $Order_{12} = Order_{21} = \emptyset$; |
| 03. **for** $l = 0$ to $r$ **do** |
| 04.     $Order_{12} = Order_{12} + r_i^1$; |
| 05.     $Order_{21} = Order_{21} + r_i^2$; |
| 06. **end for** |
| 07. $Order_{12} = Order_{12} + Order_2$; |
| 08. $Order_{21} = Order_{21} + Order_1$; |
| 09. Remove the repetitive tasks in temporary new individual $Order_{12}$; |
| 10. Remove the repetitive tasks in temporary new individual $Order_{21}$; |
| END |

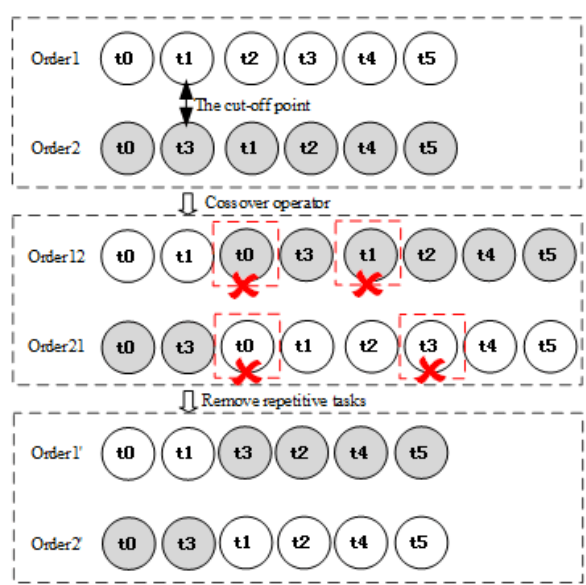

Fig. 8. The process of the single crossover of task execution order

Analogously, the crossover operators for task execution position $Loc$, encryption service level $Lev_{cf}$ and integrity service level $Lev_{ig}$ are shown as Algorithm 2. The single crossover operator of task execution position first randomly selects a cut-off point $r_1$, and then, the match area of two parent individuals of the task execution position is swapped. This is similar to that of the encryption service level, and integrity service level. An example of this operation for $Loc$, $Lev_{cf}$, $Lev_{ig}$ is given in Fig. 9.

---

**Algorithm 2: The single crossover of task execution position, encryption service level and integrity service level**

BEGIN
01. Generate at random a number $r_1, r_2, r_3, r_4 \in [0, n-1]$;
02. $Loc_{12} = Loc_{21} = \emptyset$;
03. $Lev_{au}^{12} = Lev_{au}^{21} = \emptyset$;
04. $Lev_{cf}^{12} = Lev_{cf}^{21} = \emptyset$;
05. $Lev_{ig}^{12} = Lev_{ig}^{21} = \emptyset$;
06. **for** $l = 0$ to $r_1, r_2, r_3, r_4$ do
07.     $Loc_{12} = Loc_{12} + loc_1(r_l)$;
08.     $Loc_{21} = Loc_{21} + loc_2(r_l)$;
09.     $Lev_{au}^{12} = Lev_{au}^{12} + ci_{au}^1(r_2)$;
10.     $Lev_{au}^{21} = Lev_{au}^{21} + ci_{au}^2(r_2)$;
11.     $Lev_{cf}^{12} = Lev_{cf}^{12} + ci_{cf}^1(r_3)$;
12.     $Lev_{cf}^{21} = Lev_{cf}^{21} + ci_{cf}^2(r_3)$;
13.     $Lev_{ig}^{12} = Lev_{ig}^{12} + ci_{ig}^1(r_4)$;
14.     $Lev_{ig}^{21} = Lev_{ig}^{21} + ci_{ig}^2(r_4)$;
15. **end for**
16. **for** $l = r_1 + 1, r_2 + 1, r_3 + 1, r_4 + 1$ to $n - 1$ do
17.     $Loc_{12} = Loc_{12} + loc_2(l)$;
18.     $Loc_{12} = Loc_{12} + loc_1(l)$;
19.     $Lev_{au}^{12} = Lev_{au}^{12} + ci_{au}^2(l)$;
20.     $Lev_{au}^{21} = Lev_{au}^{21} + ci_{au}^1(l)$;
21.     $Lev_{cf}^{12} = Lev_{cf}^{12} + ci_{cf}^2(l)$;
22.     $Lev_{cf}^{21} = Lev_{cf}^{21} + ci_{cf}^1(l)$;
23.     $Lev_{ig}^{12} = Lev_{ig}^{12} + ci_{ig}^2(l)$;
24.     $Lev_{ig}^{21} = Lev_{ig}^{21} + ci_{ig}^1(l)$;
25. **end for**
END

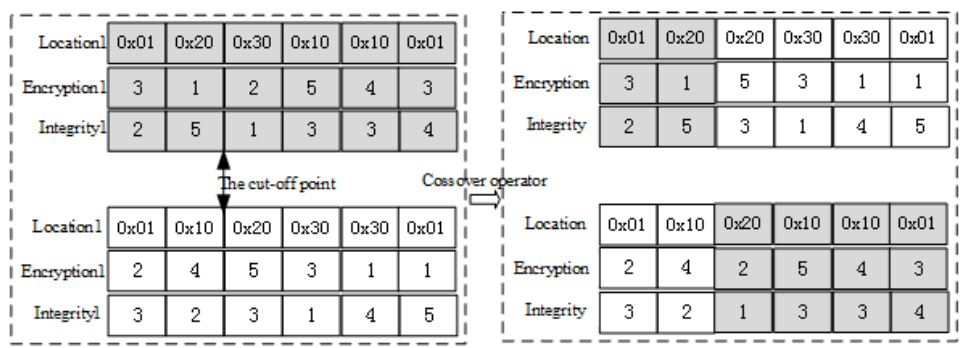

Fig. 9. The process of the single crossover of task execution position,
encryption service level and integrity service level

*5.2.3 Mutation*

The mutation operator is also a basic operator of the genetic algorithm, which plays important roles in improving the quality of the solution populations. The mutation operator is to slightly modify chromosomes to improve their fitness as well as avoid early convergence. In this section, we design the mutation operator for the set $Order$, the set $Loc$, the set $Lev_{cf}$ and the set $Lev_{ig}$, respectively. The processes of mutation operator for task execution order, the task execution location, encryption service level and integrity service level are presented in detail, respectively.

Similar to the crossover operation of task execution order, the mutation operation of task execution order must also meet with the precedence constraint. The pseudocode the mutation operation of a task execution order is given as Algorithm 3. The execution orders of the entry and end tasks are certain; therefore they can't be selected as the mutation tasks. And thus, the operator randomly chooses a mutation position $l_0 \in [1, n-2]$. Starting from task $r_0$ from an individual $Order$ of task execution order, the operator first forwards search a subset $\{r_0, \dots, r_a\}$ in which all precursors of the task $r_j$ are, and then stop the search. Then, the operator backward searches a subset $\{r_b, \dots, r_{n-1}\}$ in which all successors of the task $r_j$ are, when some task $r_b$ is reached, stop the search. At last, choose randomly a new location in the set $\{r_{a+1}, \dots, r_{b-1}\}$ for task $r_j$, and then perform insert operations. An example of the mutation of task execution order is given in Fig. 10, in which we choose randomly the task with execution sequence index 3 as the cut-off position. Then, according to the Algorithm 3, it performs the mutation operator on $Order$.

| Algorithm 3: The mutation of task execution order |
|---|
| BEGIN |
| 01. Generate at random a number $l_0 \in [1, n-2]$; |
| 02. **for** $i = 0$ to $n-1$ do |
| 03.     find some task $r_a$ which meet with the constraint $pre(r_{l_0}) \subset \{r_0, \dots, r_a\}$; |
| 04. **end for** |
| 05. **for** $i = n-1$ to 0 do |
| 06.     find some task $r_b$ which meet with the constraint $succ(r_{l_0}) \subset \{r_b, \dots, r_{n-1}\}$; |
| 07. **end for** |
| 08. Generate the set $Condidate = \{r_{a+1}, \dots, r_{b-1}\}$; |
| 09. Except the current location of the task $r_{l_0}$, choose randomly another location in the set $\{r_{a+1}, \dots, r_{b-1}\}$; |
| 11. Generate the new individual $Order = \{r_0, \dots, r_a\} + Condidate + \{r_b, \dots, r_{n-1}\}$; |
| END |

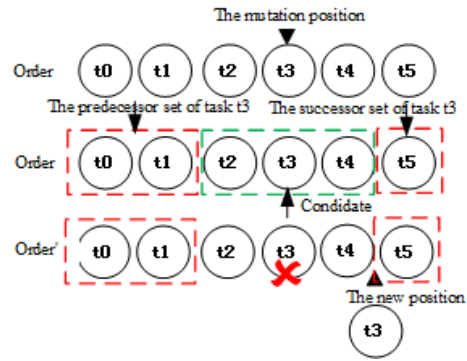

Fig. 10. The process of the mutation of task execution order

Here, the mutation operation of the task execution position, encryption service level and integrity service level are shown as Algorithm 4. They are performed by a classical operator, respectively. First, generate at random three numbers $l_1, l_2, l_3 \in [1, n-2]$ as the mutation positions of the three individuals. And then generate randomly a new valid value for the execution position, encryption service level and integrity service level, and to replace each old value corresponding to the mutation positions with a small probability. An example of this mutation operation for $Loc$, $Lev_{cf}$, $Lev_{ig}$ is given in Fig. 11.

| Algorithm 4: The mutation of task execution position |
|---|
| BEGIN |
| 01. Generate at random three numbers $l_1, l_2, l_3 \in [1, n-2]$; |
| 02. Generate at random a number $loc'(r_{l_1}) \in [0x01, 0xFF]$; |
| 03. Generate at random a number $ci'_{cf}(r_{l_2}) \in [1,5]$; |
| 04. Generate at random a number $ci'_{ig}(r_{l_3}) \in [1,5]$; |
| //Replace the $l_1$th gene value $loc(r_{l_1})$ in individual $Loc$ with $loc'(r_{l_1})$; |
| 05. $loc(r_{l_1}) = loc'(r_{l_1})$; |
| //Replace the $l_2$th gene value $ci_{cf}(r_{l_2})$ in individual $Lev_{cf}$ with $ci'_{cf}(r_{l_2})$; |
| 06. $ci_{cf}(r_{l_2}) = ci'_{cf}(r_{l_2})$; |
| //Replace the $l_3$th gene value $ci_{ig}(r_{l_3})$ in individual $Lev_{ig}$ with $ci'_{ig}(r_{l_3})$; |
| 07. $ci_{ig}(r_{l_3}) = ci'_{ig}(r_{l_3})$; |
| END |

Fig. 11. The process of the mutation of task execution position, encryption server level, integrity server level

*5.3 Initial population*

An improved genetic algorithm adopts heuristic random initialization population method to generate initial population. This method is used to schedule all tasks with the different security levels in the workflow to VMs on the APs. First, this method generates randomly the initial population, and then continue to iterate until a specified maximum number of iterations is met. The tasks execution order, task execution position, encryption service level and integrity service level are initialized as follow.

**Definition 2.** Sortable task: a task is ready if it has no any predecessor tasks, i.e. $pre(t_i) = \emptyset$; or all of its predecessor tasks have been scheduled to VMs.

Since the precedence constraint must be met between tasks in the workflow, the initialization of the task execution order is devised. First let the set $S$ to keep the sortable tasks, and choose randomly a sortable task to sort. And then choose randomly another task to sort, and continue to iterate until a feasible task order is produced. Algorithm 5 shows the pseudo-code of the initialization of tasks order.

| **Algorithm 5: The initialization of task orders** |
|---|
| BEGIN |
| 01. $S = \emptyset$; // the sortable tasks set |
| 02. $R = \{t_0\}$; //the sorted tasks set |
| 03. $T = T - \{t_0\}$; |
| 04. $r_0 = t_0$; |
| 05. $index = 0$; //the number of task sorted |
| 06. **while** $T \neq \emptyset$ do |
| 07.     **for** $t_i \in T$ do |
| 08.         **if** $pre(t_i) \subset R$ then |
| 09.             $S = S + \{t_i\}$ |
| 10.         **end if** |
| 11.     **end for** |
| 12.     choose randomly a task $t_i$ from the sortable task set $S$; |
| 13.     $++ index$; |
| 14.     $r_{index} = t_i$; |
| 15.     $T = T - \{t_i\}$; |
| 16.     $R = R + \{t_i\}$; |
| 17. **end while** |
| END |

For the initialization of task position string, generate at random a number $loc(r_i) \in [0x01, 0xFF]$ for the task $r_i$ execution position, and continue to iterate all tasks in the same way, thereby generate the set $Loc$ of the initialization of task positions. Since task $r_0$ and task $r_{n-1}$ are executed on the MD, let $loc(r_0) = 0x01$ and $loc(r_{n-1}) = 0x01$. The initialization of the encryption service level set and integrity service level set are similar to that of task position string. Algorithm 6 shows the pseudo-code of the initialization of tasks positions, encryption service level, integrity service level.

| **Algorithm 6: The initialization of task positions, encryption service level, integrity service level** |
|---|

```
BEGIN
01. loc(r₀) = 0x01;
02. loc(rₙ₋₁) = 0x01;
03. Loc = ∅, Lev_cf = ∅, Lev_ig = ∅;
04. for rᵢ ∈ T and i ≠ 0, i ≠ n − 1 do
05.     generate at random a number ps ∈ [0x01,0xFF];
06.     loc(rᵢ) = ps;
07.     Loc = Loc + loc(rᵢ);
08.     generate at random a number ci_cf(rᵢ) ∈ [1,5];
09.     Lev_cf = Lev_cf + ci_cf(rᵢ);
10.     generate at random a number ci_ig(rᵢ) ∈ [1,5];
11.     Lev_ig = Lev_ig + ci_ig(rᵢ);
12. end for
END
```

*5.4 Workflow scheduling generation*

Algorithm 7 shows the pseudocode to convert a chromosome into a schedule. For each task $t_i \in T$ of the workflow $W$, initialize its start time, end time, execution time, risk probability and transmission time to zero (step 1). For the workflow $W$, initialize its total execution energy $E(W)$, execution time $T(W)$ and risk probability $P(W)$ to zero. Step 3-19 calculate the start time $T^{ST}(t_i)$ of task $t_i$. The calculation of the start time $T^{ST}(t_i)$ can be divided into two case. The first case is that the start time $T^{ST}(t_i)$ is set 0 if task $t_i$ has no parents (step 4). The second case is that the start time $T^{ST}(t_i)$ can be computed by step 6-19 if task $t_i$ has one or more parents (step 6-19). The second case can be future subdivided into two sub-scenarios: (1) when task $t_i$ and its immediate processor task $t_r$ are executed on the same VM or APs (step 7-10), it needn't transfer the data between two different APs, and task $t_i$ can use the output data of task $t_{i-1}$ before encrypting. (2) when task $t_i$ and its immediate processor task $t_r$ are executed on different APs (step 12-18), it needs to transfer output data of task $t_r$ to its immediate successor task $t_i$ between two different APs, and it needs to employ three security services before the output data is transferred. After task $t_i$ receives the encrypted output data from all of its immediate processors' tasks, it first needs to decrypt them, and compute the sum of decryption time according to Eq. (11) (step 20). And then, based on the decryption time, step 23 compute process time of task $t_i$ according to Eq. (13). At last step 24 compute end time of task $t_i$ according to Eq. (23). With this information, the risk probability $P(W)$ of the workflow, the total execution time $T(W)$ of the workflow, and the MD's energy consumption $E(W)$ can be calculated according to Eqs. (20), (24), (25) (step 26-28). After this, the scheduling strategy corresponding to the chromosome is evaluated by the fitness value and the constraints violation. Finally, Algorithm 7 combines an improved Genetic Algorithm to produce a near optimal schedule scheme which is recorded.

| **Algorithm 7: Workflow scheduling generation** |
|---|
| BEGIN |
| 01. For each task $t_i \in T$, initialize its start time, end time, execution time, risk probability and transmission time to zero. |
| 02. **for** each schedulable task $t_i \in T$ |
| 03.     **if** task $t_i$ has no parents |

>04.    Set start time $T^{ST}(t_i) = 0$;
>05.    **else**
>06.    **for** task $t_r \in pre(t_i)$
>//task $t_i$ and its immediate processor task $t_r$ are executed on the same VM or APs
>07.       **if** $loc(t_r) = loc(t_i)$
>08.          The output data of task $t_r$ don't need be transferred to task $t_i$;
>09.          The output data of task $t_r$ don't need to be encrypted;
>10.          Obtain start time $T^{ST}(t_i) = \max\{T^{ET}(t_r) | t_r \in pre(t_i)\}$
>11.       **else**
>12.          **if** task $t_r$ isn't traversed
>13.             Compute process time of task $T^{PR}(t_r, vm_n^q)$;
>14.             Compute end time of task $T^{ET}(t_r) = T^{ST}(t_r) + T^{PR}(t_r, vm_n^q)$;
>15.             Compute the risk rate $P(t_r) = 1 - \prod_{j \in \{au, cf, ig\}} 1 - P(t_r, sl(ci_j^l))$;
>16.             Identify that task $t_r$ has been traversed;
>17.          **end if**
>18.          Obtain start time $T^{ST}(t_i) = max\{T^{ET}(t_r) | t_r \in pre(t_i)\}$
>19.       **end if**
>20. Compute the sum of decryption time of the output data of all the immediate processor tasks of task $t_i$ according to Eq. (11).
>21.    **end for**
>22. **end if**
>23.    Compute process time of task $T^{PR}(t_i, vm_n^q)$;
>24.    Compute end time of task $T^{ET}(t_i) = T^{ST}(t_i) + T^{PR}(t_i, vm_n^q)$;
>25. **end for**
>26. Calculate the total execution time $T(W)$ of the workflow according to Eq. (24);
>27. Calculate the risk probability $P(W)$ of the workflow according to Eq. (20);
>28. Calculate the MD's energy consumption $E(W)$ according to Eq. (25);
>29. Record the feasible solution $\varphi = (Order, Loc, Lev_{cf}, Lev_{ig})$.
>END

## 6. Experiments

*6.1 Experiments parameters*

In this section, to evaluate the effectiveness of SEECO strategy, we implement and simulation our strategy on Python 3.6 using a Dell R530 server configured with one CPU (2.2GHz 8 cores). we set the experimental parameters referring to the literatures [2, 56, 57]. The parameters setting is described in detail as following.

For the APs configuration, the uplink channel gain $h_{ijk}^{UL}$ is set to be equal to its downlink channel gain $h_{ijk}^{DL}$. The bandwidth among APs is set to be a constant. Each AP is configured with a VM. The computation capacities of these VMs are set to be 2.3GHz, 3.1GHz and 2.2GHz, respectively. And the processor cores are set to be 4 core, 8core and 16 core, respectively. For the mobile device, the MD's computation capacity, computational power, transmitting power, and receiving power are set as 2.36GHz, 0.5W, 0.1W and 0.05W, respectively.

For mobile service workflow, the component services and control structures are generated at random. For each component service, the input/output data and the workload follow a uniform distribution. Moreover, in order to set the proper deadline of a workflow, the minimized and maximum makespan of a workflow with the highest security service level need to be calculated. Then, the average value of the minimized and maximum makespan is set as the deadline. Thereby, the scheduling scheme can meet the risk probability and deadline constraints.

For confidentiality purpose, it provides five encryption algorithms (IDEA, DES, AES, Blowfish and RC4) to implement confidentiality service. For integrity service, it provides five hash functions (TIGER, RipeMD160, SHA-1, RipeMD128 and MD5) to implement the integrity service. The risk coefficients of these two security services are set $\lambda_{cf} = 2.5$ and $\lambda_{ig} = 1.8$.

*6.2 Impact of generic algorithm parameters*

As our strategy is based on an improved generic algorithm, we need to evaluate the impact of genetic algorithm parameters. It mainly includes four parameters, population size $pop\_size$, maximum iteration number $iterations$, crossover probability $P_c$ and mutation probability $P_m$. Four parameter configurations shown in Table 4 are used to evaluate their impacts, which is referred to [58]. The population size ranges between 10 to 1000. The maximum iteration number ranges from 50 to 500. The range of $P_c$ and $P_m$ is between 0 and 1.

Table 4. Generic Algorithm Parameters Configuration

| Configuration | $pop\_size$ | $iterations$ | $P_c$ | $P_m$ |
|---|---|---|---|---|
| Group-1 | 10-1000 | 50 | 0.2 | 0.6 |
| Group-2 | 30 | 50-500 | 0.2 | 0.6 |
| Group-3 | 30 | 100 | 0.1-0.9 | 0.6 |
| Group-4 | 30 | 100 | 0.2 | 0.1-0.9 |

Fig. 12 shows the experimental results for four groups of parameter configuration. As shown in Fig. 12(a), we observe that the MD's energy decreases gradually with the population size increasing. The reason is that the larger the population size is, the greater probability of finding optimal solutions is. However, there is no significant improvement once the population size exceeds a certain value, e.g., $pop\_size = 40$.

Fig. 12(b) shows the impact on the execution energy with the maximum number of iterations increasing. Similarly, we can also observe that the MD's energy gradually decreases with the maximum number of iteration increasing. However, the number of iterations exceeds a certain value, e.g., $iterations = 150$, the algorithm converges to the optimal solution and no significant improvement is observed.

Fig. 12(c) shows the impact on the execution energy with the mutation probability increasing. We observe from Fig. 12(c) that the lowest MD's energy can be obtained when $P_m = 0.3$. The MD's energy is unstable with $P_m$ increasing. The main reason is that high-quality chromosomes are negatively affected by the excessively large mutation probability.

Fig. 12(d) shows the impact on the execution energy with the crossover probability increasing. We can observe from Fig. 12(d) that the MD's energy decreases to a limit when $P_c = 0.5$, and then increases afterward. The main reason is that the higher the crossover probability is, the more diverse the population is. Once it exceeds a certain value, the chromosomes will become chaotic.

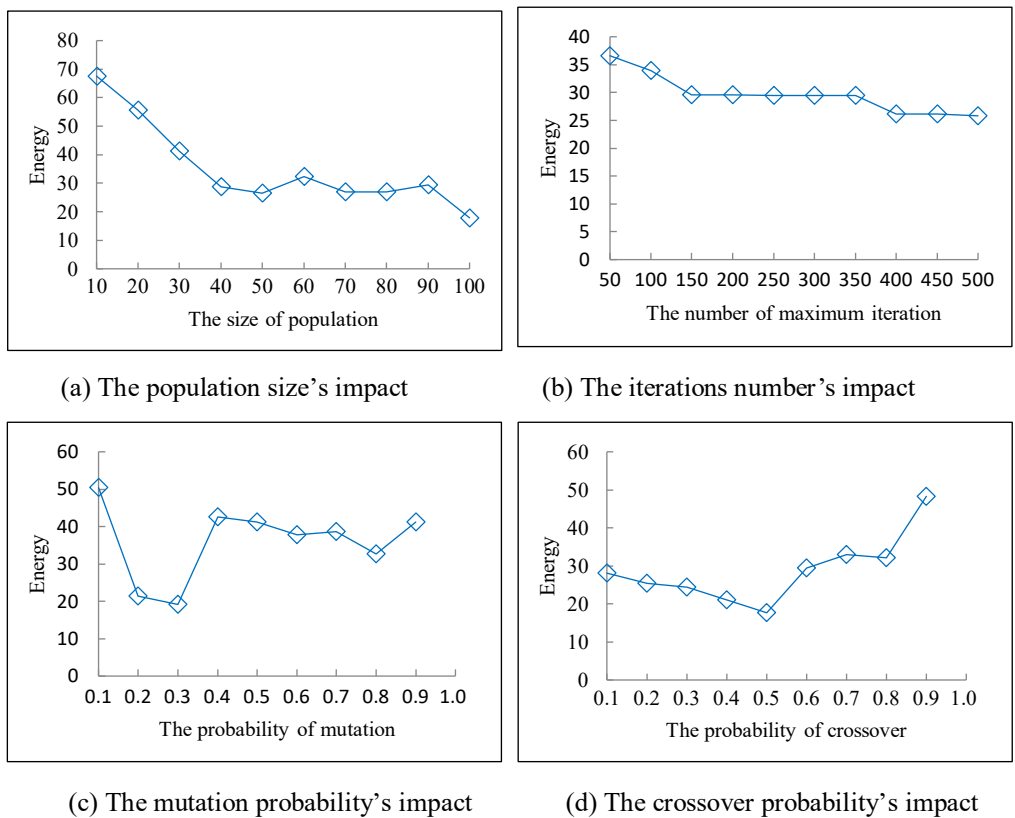

(a) The population size's impact     (b) The iterations number's impact

(c) The mutation probability's impact     (d) The crossover probability's impact

Fig. 12. The different parameters' impact

### 6.3 Comparison experiments in the execution energy

To reveal performance sensitivities, three group different experiments are conducted for 10 tasks, 30 tasks and 50 tasks of workflow, respectively. For each group experiment, the risk probability is varied from 0.1 to 1 with an increment of 0.1, and conduct these four algorithms (*Local*, *Max_Level*, *Min_Level* and *SEECO*) in terms of execution energy of workflow. The four algorithms are briefly described below:

- *Local*: This algorithm considers that all tasks of a workflow are executed on the mobile device.
- *Max_Level*: This algorithm sets all security levels of tasks on MEC equal to 1. As a result, the risk probability of each workflow is always 0.
- *Min_Level*: This algorithm doesn't incorporate any security service into tasks on the MEC. Therefore, the risk probability of each workflow is always 1.
- SEECO: This algorithm minimizes the total execution energy under the deadline and risk probability constraints in this paper.

The total execution energy obtained by the four algorithms in the experiment is shown in Fig. 13. We find that the *Local* algorithm can always get the maximum execution energy. The *Minimum Level* algorithm has the minimum execution energy. The *Max_Level* and the SEECO have moderate execution power, and the latter is superior to the former. Since the risk probabilities of both *Max_Level* and *Mini_Level* are constant, and the energy of execution is independent of the risk probabilities, the curves of both are flat. For SEECO algorithm, the energy of MD decreases rapidly with the increase of risk probability. However, when the risk probability exceeds a certain value, $P(T) = 0.5$, the energy tends to decline slowly. This lies in that the risk probability $P(W)$ is an exponential function of Eq. (20).

As the risk probability of the workflow increases, all tasks performed on the MEC require a lower level of security services, reducing the integrity of the workflow. Since the execution energy is relative to the maximum completion time of the workflow, the total execution energy eventually decreases with the increase of the risk rate. Because in the local algorithm, all the tasks are executed on the mobile device, the energy is the most. The energy of the SECCO algorithm is between the Max level and the Min level. Therefore, the SECCO algorithm can minimize the energy consumption under the risk probability and deadline constraint.

In addition, from Fig. 13, we observe that the MD's energy increases with the number of workflow tasks increasing. The least execution energy incurred by the workflow with 10 tasks, and a moderate level of execution energy incurred by the workflow with 20 tasks, and the most execution energy incurred by the workflow with 50 tasks. This lies in that the more tasks it performs, the longer it takes to execute the workflow, resulting in more execution energy.

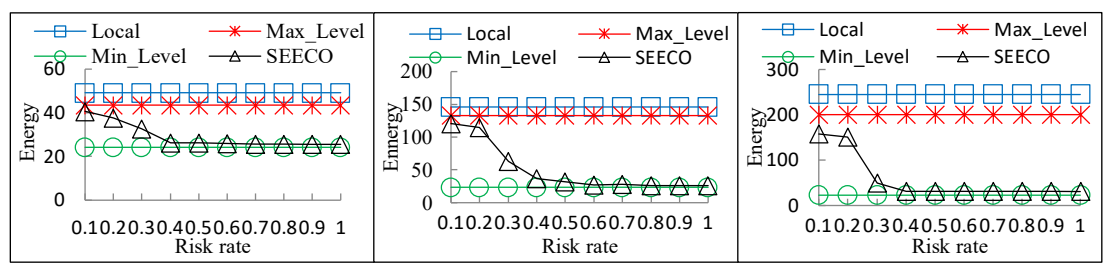

(a) The workflow with 10 tasks    (b) The workflow with 30 tasks    (c) The workflow with 50 tasks

Fig. 13. The execution energy under different risk rate constraints

*6.4 Impact of security service*

In order to evaluate the impact of the confidential service and integrity service on the execution energy, only confidentiality service and only integrity service are employed for tasks, respectively. For simplicity's sake, we use *Confi_Only* and *Integ_Only* to denote only confidentiality service and only integrity service.

Fig. 14 shows that the execution energy of *Confi_Only* and *Integ_Only* algorithms decrease with the risk probability increasing. This is because that when the risk probability of workflow increases, all the tasks executed on eNBs will demand a lower security service level. The lower security service level is, the less the makespan of workflow is, and thereby the less the execution energy is. With the same reduction of the security level, the encryption speed of *Integ_Only* decreases even faster than that of *Integ_Only*. Hence, when the increase of risk probability is equal, the execution energy of *Config_Only* decreases even faster than that of *Integ_Only*.

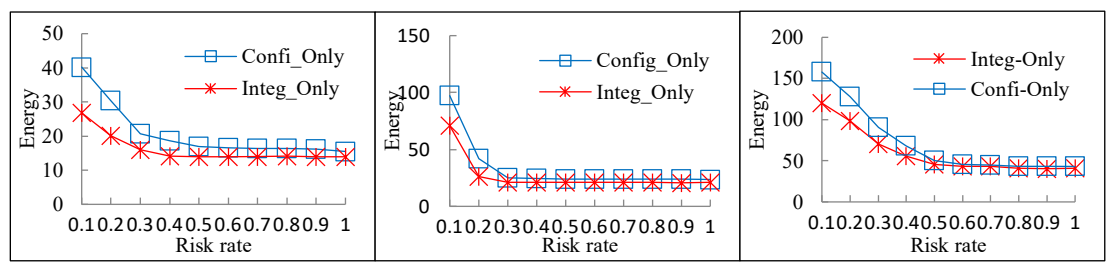

(a) The workflow with 10 tasks    (b) The workflow with 30 tasks    (c) The workflow with 50 tasks

Fig. 14. Impacts of three security services

*6.5 Impact of risk coefficient*

According to the Eqs. (18), (19), (20), the risk rate is a function of the risk coefficient. In order to evaluate the impact of the risk coefficient, we vary the risk coefficient from 0.3 to 3. Fig. 15 shows the execution energy of *Confi_Only* and *Integ_Only* with the risk coefficient varying. We observe from Fig. 15 that the execution energy of *Confi_Only* is higher than that of *Integ_Only*. The reason is that when the risk rate is constant, the security service level increases with the increase of risk coefficient according to Eq. (20), which incurs increasing the execution energy of *Confi_Only* and *Integ_Only*. What is more, when the increase of risk coefficient is equal, the execution energy of *Confi_Only* increases even faster than that of *Integ_Only*. The reason is the same to the previous section. In one word, the risk probability of workflow is almost an exponential function of risk coefficients.

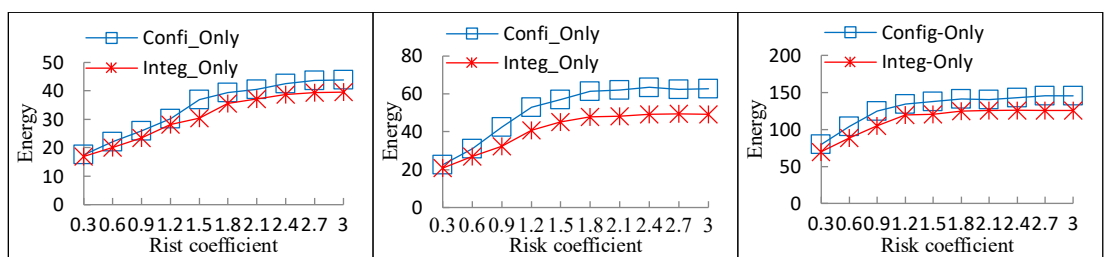

(a) The workflow with 10 tasks    (b) The workflow with 30 tasks    (c) The workflow with 50 tasks

Fig. 15. Impacts of three security coefficients

*6.6 Impact of the number of mobile edge servers*

To examine the influence of different numbers of edge servers on the execution energy, in the set of experiments, the number of edge servers are set from 0 to 10 with increments of 1. For simplicity, we use *SEECO_10*, *SEECO_30* and *SEECO_50* to represent the execution energy of SEECO for 10 tasks, 30 tasks and 50 tasks of workflow, respectively. The result reported in Fig. 16 shows that the execution energy of SEECO for three workflows decrease with the increase of the number of edge servers. The reason is that a greater number of edge servers provide more computing resource and decrease the makespan of workflow, and thereby decrease the execution energy. However, when the number of edge servers exceeds a certain value, the execution energy has no significant reduction. Therefore, for the same workflow, there is no impact on the reduction of the execution energy when the number of edge servers excessively increase.

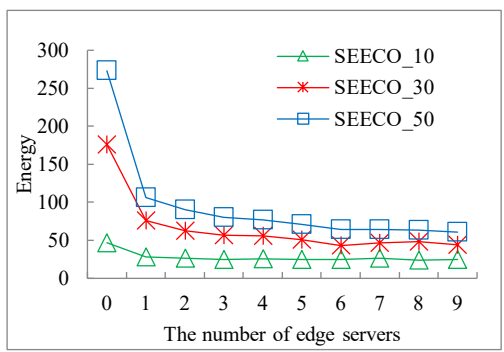

Fig. 16. Impacts of the number of edge servers

## 7. Conclusion and future work

In MEC environment, to quantify security overhead incurred by task on heterogeneous edge servers, we model a security overhead under different performance parameters, such as the CPU cores and computation frequency of MEC servers and the size of protected dataset. Based on this model, we incorporate security overheads into workflow scheduling problem, and propose a security-aware and energy-efficient workflow scheduling (SEECO) strategy. Our experimental results show that SEECO strategy can effectively decrease the MD's energy consumption while the deadline and risk rate constraints are satisfied. Especially, SEECO strategy can achieve the security guard for the security-critical tasks in MEC. In our experiment, we mainly investigate that the risk rate of security service, as well as the risk coefficient and the number of edge servers influence the execution energy of workflow. The extensive experiments using different sizes of service workflows demonstrate the effectiveness of SEECO strategy. In future work, we will study the security problem in which the workflow applications of multiple MDs can be offloaded to multiple different APs, leading to extra latency.

**ACKNOWLEDGMENTS**

This work was supported by the National Science Foundation of China (No. 61572162, 61572251, 61802095), the Zhejiang Provincial National Science Foundation of China (No. LQ17F020003), the Zhejiang Provincial Key Science and Technology Project Foundation (NO.2018C01012), and
the National Key R&D Program of China (2016YFC0800803).